\documentclass[reprint,superscriptaddress,
amsmath,amssymb,aps,prc,
floatfix,
]{revtex4-2}
\usepackage{xcolor}

\usepackage{orcidlink}
\usepackage{graphicx}
\graphicspath{{../7_bosons/HxHp_Plots/}{../8_bosons/HxHp_Plots/}{../hProbvN_Plot_Hp/}{../I8_Prob_Density_N8_J0_GSEnergies_Hp/}{../Q_N8_J0gs_Overlaps_Plot_w2VLs_c09/}{../iVenn_Diagram_N4N8N12_J0gs/}{../iVenn_Diagram_N6N9N12_J0gs/}{../iClusterNum_Dist_N12J0_C2b/}{../iClusterData_N9N12_9_Hp/}{../iClusterData_N9N12_12_Hp/}{../iClusterNum_Dist_N12J0_C3b_overlay/}{../iClusterData_N8N12_12_Hp/}{../iClusterData_N8N12_8_Hp/}{../MagDist_Plot/}}
\usepackage{amssymb}
\usepackage{makecell}
\usepackage{lipsum}

\usepackage{xr-hyper}
\usepackage{hyperref}
\hypersetup{breaklinks=true,colorlinks=true,linkcolor=blue,citecolor=blue,filecolor=magenta,urlcolor=cyan}

\begin{document}
\title{Are the ground states of randomly interacting bosons random?}
\author{Charles White\,\orcidlink{0000-0002-1962-7300}}
\author{Alexander Volya\,\orcidlink{0000-0002-1765-6466}}
\affiliation{Department of Physics, Florida State University, Tallahassee, Florida 32306, USA}
\author{Declan Mulhall\,\orcidlink{0000-0001-8379-0148}}
\affiliation{Department of Physics and Engineering, University of Scranton, Scranton, Pennsylvania 18510-4642, USA}
\author{Vladimir Zelevinsky}
\affiliation{Department of Physics and Astronomy and Facility for Rare Isotope Beams, \\
Michigan State University, East Lansing, MI 48824-1321, USA}
\date{\today}
\begin{abstract}

Bosonic degrees of freedom and their emergence as a part of complex quantum many-body dynamics,
symmetries, collective behavior, clustering and phase transitions play an important role in modern studies of quantum systems.
In this work we present a systematic study of many-boson
systems governed by random interactions.  Our findings show that ground states of randomly interacting bosons are not random, being dominated by a few collective configurations containing condensates of clusters.
\end{abstract}

\maketitle


\section{Introduction}

Quantum many-body systems display a variety of remarkable phenomena that emerge as a result of simple interactions among the constituents.
Bosonic degrees of freedom and their interactions play an important part in the emergent phenomena. In this work we address the generic features exhibited by randomly interacting systems of bosons.
The revolutionary ideas about statistical properties of quantum systems date back to the Gaussian Orthogonal Ensemble (GOE) of random matrices introduced by Wigner
\cite{wigner:1955,wigner:1957,wigner:1958}, who proposed that a random matrix ensemble invariant with respect to the choice of basis be used to describe complex, 
or chaotic in modern terminology, quantum systems.
Random matrix theory is a powerful and widely used modern-day tool; for overview  see Refs. \cite{guhr:1998,weidenmuller:2009,francesco:2015}, and references therein. 
The well-known physics of many realistic systems, especially at relatively low excitation energies,
suggest that low-rank embedded ensembles \cite{kota:2001} and, in particular, two-body random ensembles (TBRE) are quite relevant 
\cite{bohigas:1971,brody:1981,french:1970,french:1971,zelevinsky:1993a}. 
For numerous realistic atomic and nuclear systems \cite{flambaum:1994,weidenmuller:2009,zelevinsky:1996a}, high level density leads to strong mixing 
of basis states, and the observable properties  approach those given by the predictions of random matrix ensembles.

%

Over the last two decades the question of the emergence of regularities out of chaos  has been a particularly interesting and 
rapidly developing topic. One of the main highlights is an observation of an unusually high likelihood for the ground 
state of a random system to have zero spin \cite{johnson:1998}. 
This remains true  even if the two-body interaction does not 
favor pairing or any other obvious force favorable to a
net-zero angular momentum arrangement.  
Multiple further investigations have followed  \cite{johnson:1998,zelevinsky:2016:art,abramkina:2011:art,zelevinsky:2004:art,mulhall:2000:art,chau-huu-tai:2002,zhao:2002,papenbrock:2004,zhao:2004,
johnson:2007,weidenmuller:2009}
and many more remarkable features that emerge out of chaos have been found.
In fermionic systems, the emergence of rotational and vibrational regularities \cite{abramkina:2011:art,johnson:2007} naturally
connects this topic to the interacting boson model \cite{arima:1981a} and to the more in-depth questions of higher symmetries \cite{bijker:2000a,chau-huu-tai:2002,bijker:2002}
known to be intrinsic to bosonic systems and dynamics of shape and phase transitions  \cite{cejnar:2009}. 

While bosonic systems have been discussed for a while  \cite{zhao:2003,chau-huu-tai:2002}, 
there has been a resurgence 
of interest \cite{lu:2015,fu:2018,zhao:2018} because of the importance that bosons play in our general understanding of emergent phenomena, such as  symmetries, formation of effective degrees of freedom, clustering, and collective dynamics that include pairing and phase transitions. 

In this work we study bosonic systems with two-body random interactions. We limit the scope to systems of identical bosons, although our results are likely to be valid more generally.  We also limit our consideration to ground states. Other properties, such as emergence of bands and transitions, will be discussed 
elsewhere, see Ref. \cite{white:2023}. 

\section{Bosonic Random ensemble \label{sec:BE}}
We consider a system of $N$ identical bosons with integer spin $\ell$ that interact via the most general two-body Hamiltonian 
\begin{equation}
H=\sum_{L=0,2,4\dots}^{2\ell} V_L \sum_M P^\dagger_{LM} P_{LM}.
\label{eq:1}
\end{equation}
The $P^\dagger_{LM}$ and $P_{LM}$ are boson pair creation and annihilation operators with angular momentum $L$ and magnetic projection $M.$ 
We normalize them so that a pair state created from the vacuum by any of the operators $P^{\dagger}_{LM}$ is normalized to one.
Since the bosons are identical, they can only be in a symmetric state, which requires $L$ to be even.  

How the many-boson states can be classified in terms of the rotational group 
and, specifically for a system of $N$ bosons with spin $\ell$, how many states there are with total spin $J,$ labeled as $D_{\ell N}(J),$ 
are important and non-trivial questions. For completeness of this presentation, we review this subject in Appendix \ref{apx:A}; 
additional information can be
found in the textbook \cite{zelevinsky:2017}. 
In certain cases, $D_{\ell N}(J)$ is known analytically; generally, as a function of $J$ it is a peaked curve, so that the number of 
states with low and high spin is small,  see Fig.~\ref{fig:disbos}. Cases of unique states $D_{\ell N}(J)=1$, which include the 
aligned state of maximum spin $J_{\rm max}=\ell N$,
are special as in these cases the energy is a linear function of the interaction parameters $V_L.$ 

For the two-body  random forces, we assume an ensemble of Hamiltonians where the matrix elements $V_{L}$ of the two-body interaction 
are selected at random from a normal distribution so that
\begin{equation}
\overline{V_L}=0 \quad {\rm and} \quad \overline{V^2_L}=1.
\label{eq:ens}
\end{equation}

There are also the well-known special two-body Hamiltonians: 
pairing, monopole, and square of the total angular momentum operator, which we review in more detail in Appendix  \ref{apx:B}, see also 
Ref.~\cite{zelevinsky:2017}.  The monopole operator, which is related to the number of particles, and the angular momentum operator squared
both commute with any Hamiltonian; thus, we remove these collective components from each realization of Hamiltonian \eqref{eq:1}, resulting in 
the slightly modified Hamiltonian $H'$ of the ensemble, which we refer to collectively as the ``primed'' ensemble.
The exact procedure is discussed in Appendix \ref{apx:prime}.
Clearly, the wave functions do not change; the monopole term only shifts the overall energy, while the $J^2$ term gives a contribution proportional to 
$J(J+1)$ to the energy of each state. 
We found no significant difference in the results that are of interest from the removal of the two collective components; 
however, doing this helps to address a potential concern related to these collective terms dominating the statistics, see Ref. \cite{white:2023}.

For each random realization of the Hamiltonian,  we study its ground state wave function, energy, spin and other properties for systems with different numbers of particles. We also explore correlations and multi-boson transitions between the ground states of different-sized systems. 
Numerical results for various systems that we discuss throughout this work are shown in Table \ref{tab:main}. The first three columns identify the boson spin $\ell$,
number of bosons $N$, and the spin of interest $J,$ the fourth column shows the number of states with this spin in the system  $D_{\ell N}(J), $ 
and the fifth column gives the total number of spin states $D=\sum_J D_{\ell N}(J).$
We discuss other columns as we continue with our presentation. 

\section{Ground state spin probability distribution\label{sec:gs}}
\begin{figure}[h]
\begin{center}
\includegraphics[width=0.9\linewidth]{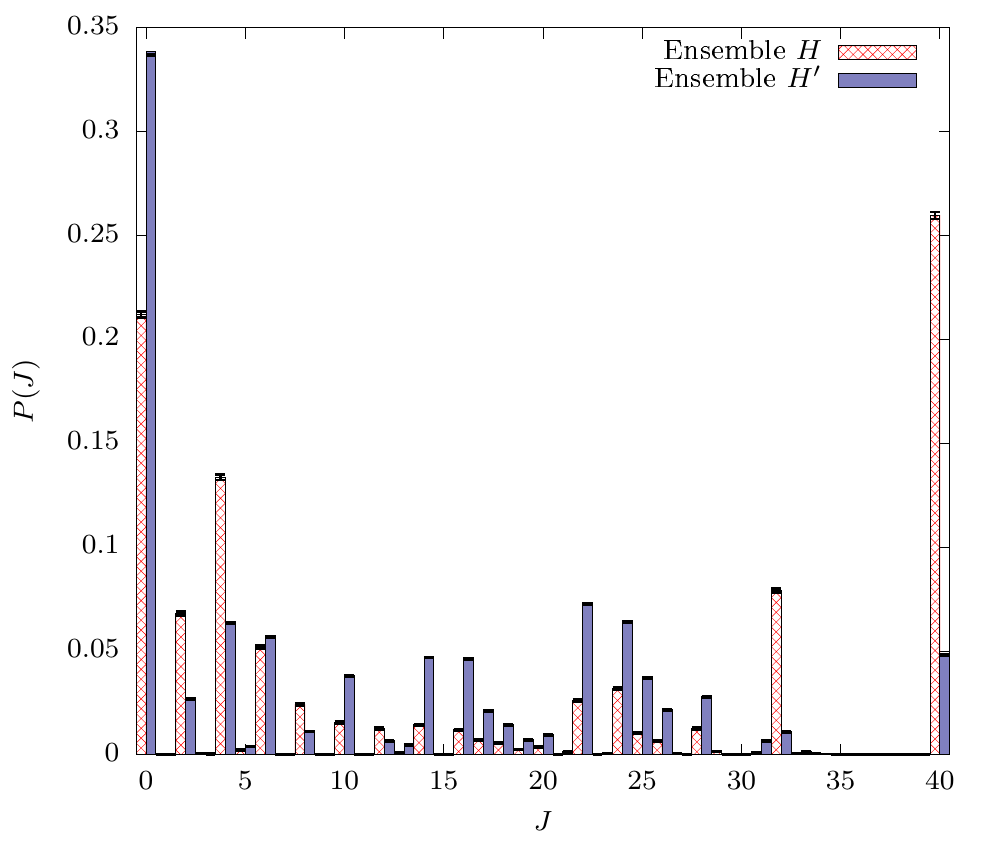}

\end{center}
\caption{\label{fig:dis8} Comparison of ground state spin distribution for the original and primed ensembles.
System of $N=8$ bosons with spin $\ell=5$ is considered. 
}
\end{figure}

\begin{figure}[h]
\begin{center}
\includegraphics[width=0.9\linewidth]{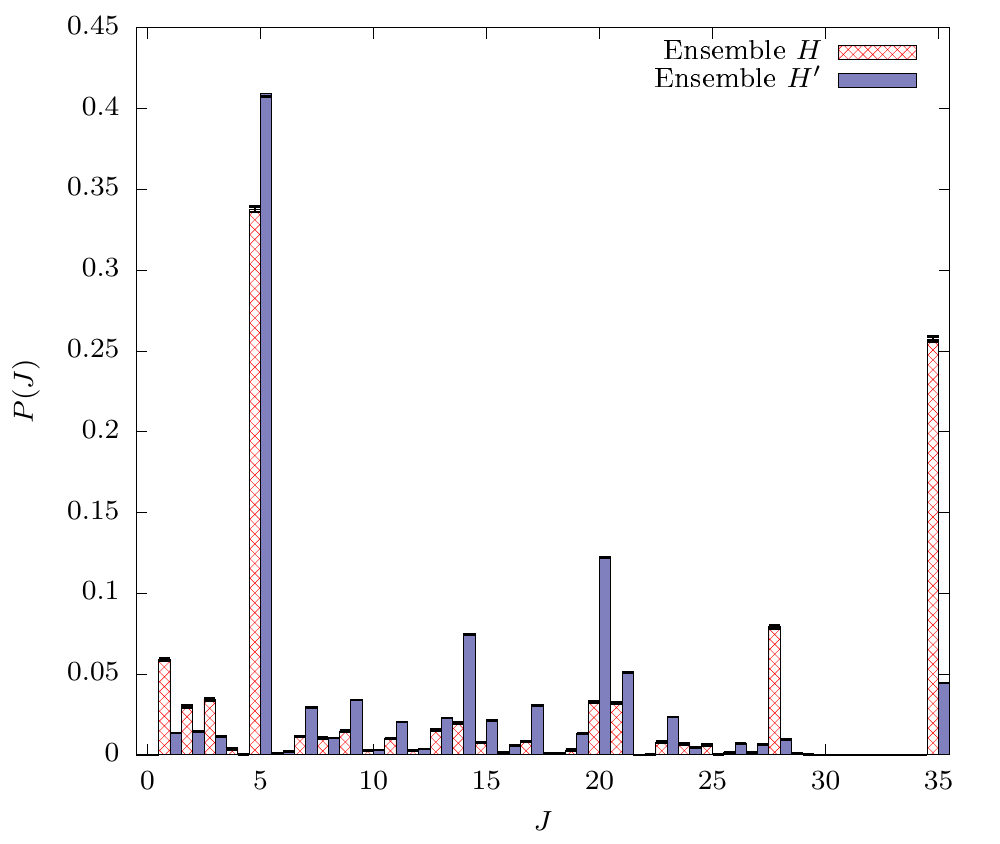}
\end{center}
\caption{\label{fig:dis7} Comparison of ground state spin distribution for the original and primed ensembles.
System of $N=7$ bosons with spin $\ell=5$ is considered. 
}
\end{figure}

We begin with a discussion of the ground state spin distribution in Figs. \ref{fig:dis8} and \ref{fig:dis7}, which show the probability $P(J)$ of 
the ground state having spin $J$ in  8 and 7 particle systems of $\ell=5$ bosons, respectively. 
This is a known result, discussed previously by many authors, and is surprisingly similar to the systematics found in fermionic systems.  
Following are some of the key observations. 
In a large number of cases, typically between  30 to 50\%, the ground state spin is equal to either zero for an even number of bosons or to $\ell$ (the spin of an individual boson) if the number of bosons is odd. 
In disproportionately  many cases, the ground state has the maximum angular momentum possible $J_{\rm max}=N\ell.$ 
The probability of $J=J_{\rm max}-N=(\ell-1)N$ is also enhanced. 
The results for $P(J)$   from numerical studies using the ``primed'' ensemble are shown in the 6th column of Table \ref{tab:main}.

\begin{figure}[h]
\begin{center}
\includegraphics[width=0.9\linewidth]{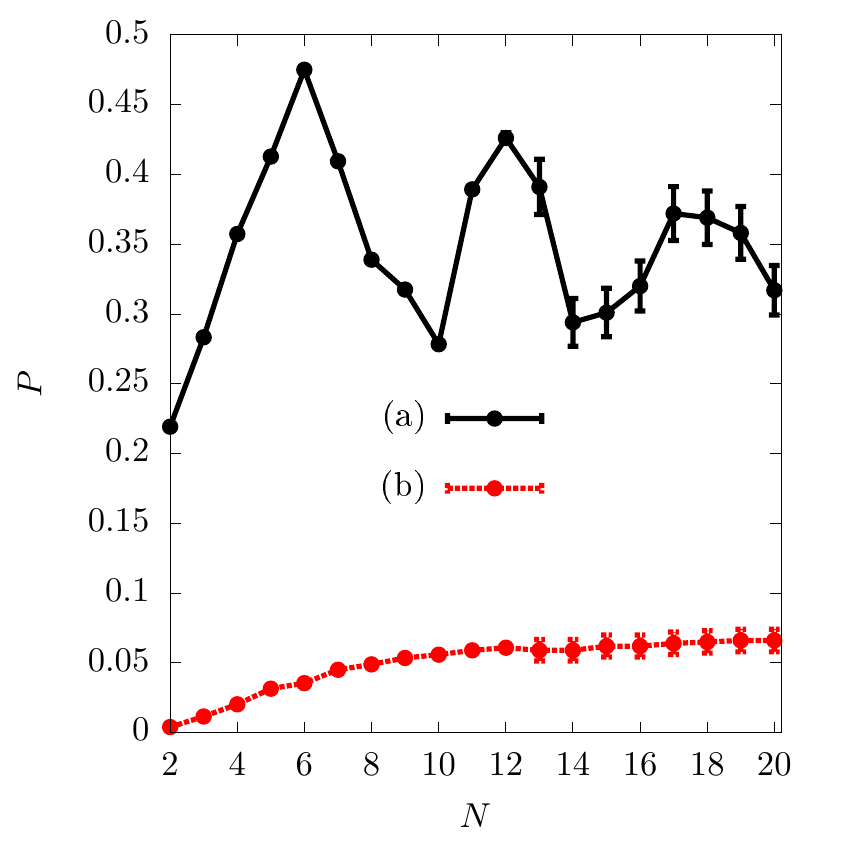}
\end{center}
\caption{\label{fig:4} For bosons with spin $\ell=5$, the probability of observing a ground state of a certain
spin $J$ is shown as a function of boson number $N.$
Line (a) shows the probability of the ground state spin  $J=0$ for even $N$ and $J=\ell$ for odd $N.$ Line (b) shows the probability of 
$J_{\rm max}=N \ell.$
To expedite our studies for systems with $N > 12$,  ensembles of smaller size were used which resulted in noticeable statistical error shown with the error bars.
}
\end{figure}

The likelihood  of the ground state being aligned with a maximal spin is affected by the collective $J^2$ term;  removal of this term reduces $P(J_{\rm max}).$ 
The maximally aligned state represents a condensate of aligned bosons with energy fully determined by a single matrix element $V_{2\ell},$
\begin{equation}
E=\frac{1}{2} N(N-1) V_{2\ell}.
\label{eq:jmax}
\end{equation}
Being collectively scaled with the number of particles and dependent on this single matrix element, the enhancement $P(J_{\rm max})$ is not unexpected, 
although there is no answer for an exact probability. Some assessments based on extreme eigenvalue deviations can be made \cite{dean:2006}.
Other high spin states with enhanced chances to appear in the ground state are also structurally very simple and generally 
depend only on the few matrix elements $V_L$ with largest $L.$
This  is evident from the comparison of the original and primed ensembles  in 
Figs. \ref{fig:dis8} and \ref{fig:dis7}.
The behavior of $P(0), P(\ell), $ and $P(J_{\rm max})$ for a system of bosons with spin $\ell=5$ as a function of the boson number $N$ is shown in 
Fig. \ref{fig:4}.


%

\section{Example of $d$-bosons\label{sec:dbos}}

As an introductory discussion, in this section we present the analytically solvable case of $\ell=2,$ $d$-bosons.
This problem is analytically solvable thanks to the additional symmetry that brings in a conserved quantum number.
Details of bosonic algebras can be found in a number of references
\cite{iachello:1987,casten:1993,frank:2009,zelevinsky:2017}.
The spin statistics of ground states for a $d$-boson system can be determined analytically \cite{chau-huu-tai:2002,lu:2015} because
there are enough quantum numbers to make the eigenvalues of the Hamiltonian
linear functions of the two-body interaction parameters $V_L,$ eq.~(\ref{eq:23}). 
This is an instructive example that helps to guide our discussion that follows. 

For a given $N$, the number of unpaired particles (called seniority) can be $\nu=N,\ N-2,\ N-4,\dots ,$ with the smallest value being 0 or 1 if the number of particles is, correspondingly, even or odd.
Any Hamiltonian for
$d$-bosons commutes with the pairing Hamiltonian making seniority $\nu$ a good quantum number. 
Then, among $\nu$ unpaired particles, we can have some triplets that are coupled to zero angular momentum; thus, the number $f$ of ``free'' particles that are neither in pairs nor in triplets is 
$f=\nu,\ \nu-3,\ \nu-6,\dots$, with the smallest number being the remainder from division of $\nu$ by 3 (modulo). 
These uncoupled $f$ particles are the ones producing the angular momentum 
\begin{equation}
J=2f,\ 2f-2,\ 2f-3,\ 2f-4,\dots,\ f,
\label{eq:28}
\end{equation}
which can take all integer values between $2f$ and $f$ with the exception of $2f-1.$

Thanks to seniority being conserved, all states can be uniquely identified by their spin and seniority. Thus, the energy from Hamiltonian \eqref{eq:1} is a linear function of  the three interaction parameters $V_0,$ $V_2,$ and $V_4.$ 
The energy for $N$ $d$-bosons as a function of seniority and angular momentum is 
\begin{equation}
E(\nu,J)=-\beta \nu(\nu+3) + \gamma J(J+1),
\label{eq:29}
\end{equation}
where the coefficients $\beta$ and $\gamma$ are given in terms of the two-body matrix elements
\begin{equation}
\beta=\frac{1}{10} V_0 - \frac{1}{7} V_2 +  \frac{3}{70} V_4,
\label{eq:30}
\end{equation}
\begin{equation}
\gamma=\frac{1}{14}( V_4-V_2).
\label{eq:31}
\end{equation}
Here we disregard some constant terms that depend only on the number of particles. With only three interaction 
parameters, any Hamiltonian can always be written as a linear combination of pairing, monopole, and angular momentum squared terms, 
which provides an alternative perspective on the result. See Appendix \ref{apx:B}.

Assuming that the parameters $V_L$ obey a normal distribution, the joint probability distribution for $\beta$ and $\gamma$ is
\begin{equation}
P(\beta,\gamma)=\frac{70}{\sqrt{3} \pi }\exp\left({-\frac{4}{3} \left(25 \beta^2-65 \beta \gamma+79 \gamma^2\right)}\right).
\label{eq:32}
\end{equation} 
From this we can find various probabilities; for example, the  probability for the case where $\beta$ and $\gamma$ are both positive (same for both negative) is
\begin{equation}
P(\beta>0,\gamma>0)=\frac{1}{4}+\frac{1}{2\pi} \arctan \left( \frac{13}{7 \sqrt{3}}\right)\approx 0.38.
\label{eq:33}
\end{equation}


These rules provide a direct strategy for determining $P(J).$ We will not go into the details 
of the analytical analysis built around eq.~(\ref{eq:32}), the results can be found in Ref. \cite{lu:2015}.
Because of pair and triplet clusters, the systematics have a periodicity of 6 in the number of particles. This periodicity becomes exact in the asymptotic limit of large $N.$ The ground state spin can only be  $J=0$, $J=2$, or $J=2N$; the asymptotic probabilities $P(J)$ over the period are summarized in Tab. \ref{tab:pp}.

\begin{table}[h]
\begin{tabular}{|c|ccc|}
\hline
$N$& $P(0)$  & $P(2)$  & $P(J_{\rm max})$ \\
\hline
$6k$ & 57 & 0 & 43\\
$6k\pm 1$ & 2 & 55 & 43 \\
$6k\pm 2$ & 19 & 38 & 43 \\
$6k\pm 3$ & 40 & 17 & 43 \\
\hline
\end{tabular}
\caption{\label{tab:pp} Table shows asymptotic probabilities ($N\gg1$), which have a periodicity of 6 in the particle number, 
to see ground state of spin $J=0\,$ $J=\ell=2\,$ and $J=J_{\rm max}=2N$ for $d$-bosons. Probabilities are expressed in percentage. 
Integer $k$ denotes the period. 
}
\end{table}


To summarize this analytic example,
the ground states of $d$-boson systems  are combined from spin-zero pairs and triplets or are condensates of aligned bosons
in the case of $J_{\rm max}.$
Because interactions are of the two-body type, the triplets of spin zero do not explicitly contribute to the energy
and, depending on if pairing interaction is attractive or
repulsive, the seniority is correspondingly minimized or maximized.
This clustering into pairs and triplets leads to periodicity of 6 in the results,  as shown in Tab. \ref{tab:pp}.

\section{Ground-state energy distribution\label{sec:energy}}
There are situations when only one state with given quantum numbers exists in the system. This includes 
the previously discussed example of $d$-bosons (with seniority providing an additional quantum number), 
cases of an aligned state with $J_{\rm max}$, and many examples discussed in Appendix \ref{apx:A}
where $D_{\ell, N}=1;$ we specifically mention there $J=0$ triplets of bosons with even $\ell$, which are unique ($D_{\ell, N=3}=1$ for even $\ell$), see sec. \ref{apx:A2}.

In these cases, the wave function $|\psi\rangle$ is uniquely determined by the state's quantum numbers, which makes the energy 
\begin{equation}
E=\sum_L c_L V_L,\,\,{\rm where}\quad c_L= \sum_M \langle\psi| P^\dagger_{LM} P_{LM} |\psi\rangle,
\label{eq:23}
\end{equation}
a linear function of the interaction parameters $V_L.$  
Some discussion of analytically solvable models based on linearity is found in Ref. \cite{chau-huu-tai:2002}.

From the definition in eq.~(\ref{eq:23}), it is clear that 
\begin{equation}
c_L\ge 0.
\label{eq:24}
\end{equation}
It is also clear that 
\begin{equation}
\sum_L c_L=\frac{N(N-1)}{2}=\sigma_{\rm max},
\label{eq:25}
\end{equation}
since, similar to the monopole interaction,  Appendix \ref{apx:B1}, this sum counts the total number of pairs in the state. 
We label this sum in eq.~\eqref{eq:25} as $\sigma_{\rm max}$ because in the ensemble \eqref{eq:ens} 
\begin{equation}
\overline{E}=0\,,\quad \overline{E^2}=\sum_L c_L^2,
\label{eq:26}
\end{equation}
and conditions (\ref{eq:25}) and (\ref{eq:24}) constrain the width of the energy distribution
\begin{equation}
\frac{\sigma^2_{\rm max}}{\ell+1} \le \overline{E^2}\le \sigma^2_{\rm max}.
\label{eq:27}
\end{equation}
The minimum is realized when all $\ell+1$ coefficients $c_L$ are the same. The maximum, which is  the most relevant limit for us,  is realized for a pair condensate when only one $c_L$ is non-zero.  Thus, $\sigma^2_{\rm max}$ is the maximal energy variance that is possible for a 
state whose energy is a linear function of interaction parameters  \eqref{eq:23} in the two-body random ensemble \eqref{eq:ens}.
Formation of a condensate that leads to a broad energy distribution reaching  $\sigma_{\rm max}$ explains the preponderance of  $J_{\rm max}$ being a ground state, as well as the enhanced chances of some other states with high spin.  

\begin{figure}[h]
\begin{center}
\includegraphics[width=0.9\linewidth]{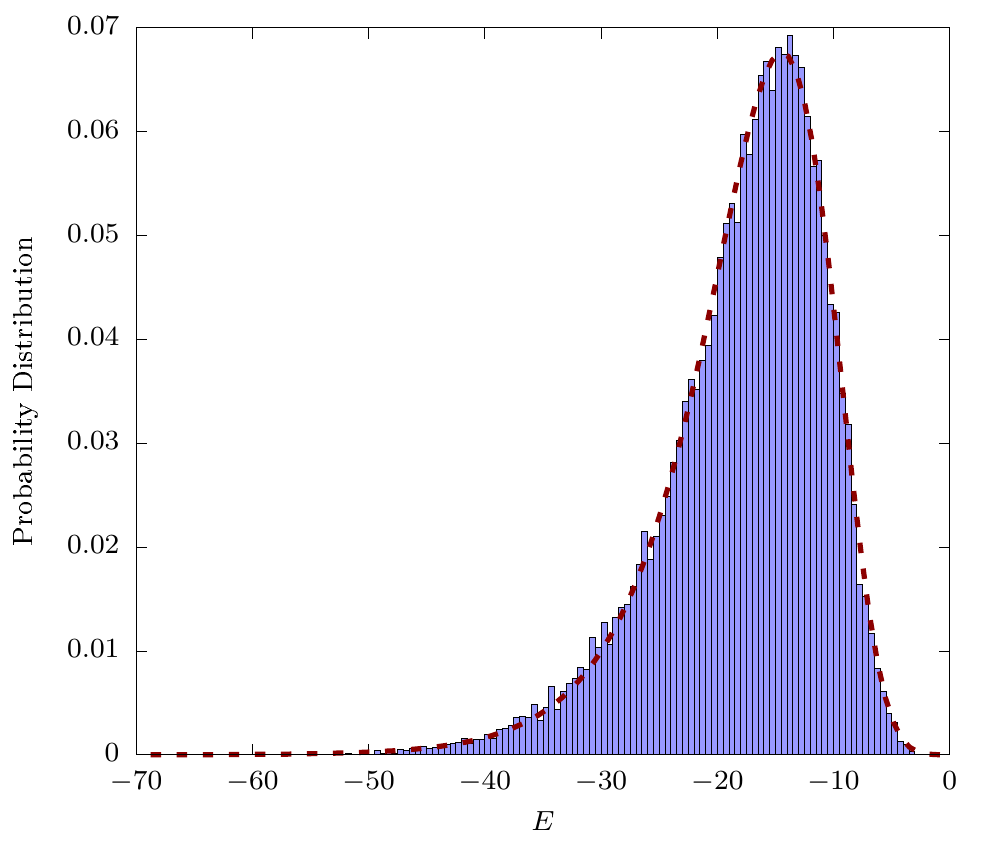}
\end{center}
\caption{Distribution of $J=0$ ground state energies in a system with $N=8$ and  $\ell=8.$ The distribution is compared 
with the Gumbel distribution in eq. (\ref{eq:gumbel}), where $a=-14.46(4)$ and $b=0.183(2)$
\label{fig:gumbel}  
}
\end{figure}

The energies of the ground states themselves are independent identically distributed variables; thus, their 
distribution follows one of three universal distributions \cite{hansen:2020,jan-beirlant:2004,galambos:1987,gumbel:1958}. Here we have a case of the Gumbel distribution 
\begin{equation}
G(E)={b}\,\exp\left [{b(E-a)}- \exp\left ({b(E-a)}\right ) \right ]
\label{eq:gumbel}
\end{equation}
because our random ensemble is 
given by interaction matrix elements with a normal distribution, and thus the probabilities of extreme values of energy fall faster than any power law. 
The numeric results are in very good agreement with eq. (\ref{eq:gumbel}). An example of the ground state energy distribution is 
compared with the Gumbel function in Fig. \ref{fig:gumbel}.  
The parameters of the Gumbel distribution can be associated with the number of degrees of freedom \cite{hansen:2020,kota:2018,palassini:2008}. 

For a Gaussian distribution, which is the case  for eq. (\ref{eq:23}) at the start of this discussion, the parameters of the distribution $a$ and $b$ depend 
on the size of the set of normally distributed random numbers from which the minimum or maximum is picked. We associate it with some effective
dimensionality $\mathcal{D}$ of states competing to appear as ground states.  
\begin{equation}
|a| = \sqrt{2 \overline{E^2}}\,\,{\rm erf}^{-1}\left(\frac{{\cal D}-2}{{\cal D}}\right), \quad b = {\cal D} f(a)
\label{eq:ab}
\end{equation}
where ${\rm erf}^{-1}$ denotes the inverse error function, and function $f$ is the normal distribution of energies $E$
centered at zero and with variance $\overline{E^2}.$ We discuss the minimum, thus $a$ is negative.

Using the Gumbel distribution parameters $a$ and $b$ observed numerically and
solving these equations allows us to determine the effective dimensionality $\mathcal{D}$ and the
variance $\overline{E^2}.$
In the limit of large $\mathcal{D}$, the expressions \eqref{eq:ab} can be simplified using the product logarithm function, see  
Ref.~\cite{hansen:2020}.
In Table \ref{tab:main}, we include columns that, for each system, show the effective dimension $\mathcal{D}$ and the variance relative to its maximal value 
\eqref{eq:27}, namely  $\sqrt{\overline{E^2}}/\sigma_{\rm max}.$ This is done by solving \eqref{eq:ab} given $a$ and $b$ from a numerical fit, such as the one shown in
Fig. \ref{fig:gumbel}.
Our studies show that due to the small dimensionality of $\mathcal{D}$,
this procedure overestimates $\mathcal{D}$  and  slightly underestimates $\sqrt{\overline{E^2}}$.  
For example, numerical studies of normal distributions that take as input unit width,$\sqrt{\overline{E^2}}=1,$ and ${\cal D}=5,6,10,100$ 
result in the corresponding list of $(\mathcal{D},\sqrt{\overline{E^2}})$ being (6.5, 0.93),  (7.5, 0.93)  (12, 0.97) (106, 1); this gives an idea about the level of error 
in the inversion procedure. 
The two-body random ensemble is certainly more complicated, which is evident
from the case of $N=3,$ $\ell=6$ for $J=0$, where it is analytically known from eq \eqref{eq:13a} that $\sigma_{\rm max}=\sqrt{\overline{E^2}}=3,$ yet our procedure 
gives $\sqrt{\overline{E^2}}\approx2;$ as seen  in Table \ref{tab:main} $\sqrt{\overline{E^2}}/\sigma_{\rm max}=0.67$.  Despite this, the results are certainly reflective of the width of the energy 
distribution and the number of states competing for the 
ground state position. 
%

The following conclusions can be drawn from this discussion. First, in all cases studied, only about a 
dozen states are competing for the ground state position, as evident from the values of $\mathcal{D}$ inferred from the distribution of ground state energies. 
Compare $\mathcal{D}$  with $D_{\ell N}$ in  Table \ref{tab:main}.
Second, 
the width of the ground state energy distribution is a substantial fraction of the maximum allowed value that occurs in condensates, 
indicating that only a few two-body matrix elements are responsible for the ground state structure. 

\begin{widetext}

\begin{table}[htb]
    \centering
    \begin{tabular}{@{}|cccccc||cccccccc||cccc|@{}}
    \hline
        $\ell$ & $N$ & $J$ & $D_{\ell N}(J)$ &$D_{\ell N}$ &   $P(J)$ [\%] & $q_1$ & $q_2$ & $q_3$ & $q_4$ & $q_4$ & $q_6$ & $q_7$ & $D_{\rm gs}$ & $a$ & $b$ & $\mathcal{D}$ & $\sqrt{\overline{E^2}}/\sigma_{\rm max}$   \\
        \hline
        3 & 8 & 0 & 4 & 151 & 63.9 & 0.549 & 0.356 & 0.081 &  &  &  &  & 2.6 & -8.49 & 0.162 & 5.7 & 0.33 \\
        4 & 8 & 0 & 7 & 526 & 39.8 & 0.562 & 0.161 & 0.087 &  &  &  &  & 2.6 & -9.2 & 0.182 & 7 & 0.31 \\
        4 & 12 & 0 & 20 & 3788 & 64.2 & 0.391 & 0.339 & 0.172 & 0.049 &  &  &  & 4 & -20.92 & 0.085 & 7.4 & 0.29 \\
        5 & 4 & 0 & 2 & 55 & 35.7 & 0.687 & 0.313 &  &  &  &  &  & 1.9 & -3.64 & 0.599 & 9.6 & 0.48 \\
        5 & 5 & 5 & 10 & 141 & 41.3 & 0.497 & 0.194 & 0.177 & 0.053 &  &  &  & 4.2 & -4.65 & 0.459 & 9.3 & 0.38 \\
        5 & 6 & 0 & 6 & 338 & 47.5 & 0.395 & 0.291 & 0.156 & 0.114 &  &  &  & 4.2 & -6.6 & 0.321 & 9.2 & 0.36 \\
        5 & 7 & 5 & 34 & 734 & 40.9 & 0.482 & 0.177 & 0.166 & 0.078 & 0.051 &  &  & 4.7 & -8.63 & 0.24 & 8.9 & 0.34 \\
        5 & 8 & 0 & 12 & 1514 & 33.9 & 0.6 & 0.218 & 0.127 &  &  &  &  & 3.1 & -11.67 & 0.174 & 8.7 & 0.35 \\
        5 & 8 & 32 & 7 & 1514 & 1.1 & 0.513 & 0.486 &  &  &  &  &  & 2 & -11.1 & 0.204 & 10.1 & 0.31 \\
        5 & 12 & 0 & 52 & 16660 & 42.6 & 0.45 & 0.173 & 0.146 & 0.1 & 0.053 & 0.03 &  & 5.4 & -23.21 & 0.084 & 8.3 & 0.3 \\
        6 & 3 & 0 & 1 & 25 & 22.1 & 1 &  &  &  &  &  &  & 1 & -2.566 & 0.859 & 9.7 & 0.67 \\
        6 & 4 & 0 & 3 & 86 & 49.1 & 0.457 & 0.381 & 0.162 &  &  &  &  & 2.8 & -4.02 & 0.61 & 11.3 & 0.48 \\
        6 & 6 & 0 & 8 & 676 & 46.9 & 0.384 & 0.27 & 0.181 & 0.094 &  &  &  & 4.6 & -8.26 & 0.287 & 10.8 & 0.41 \\
        6 & 7 & 6 & 63 & 1656 & 46.1 & 0.337 & 0.216 & 0.196 & 0.106 & 0.061 & 0.02 &  & 6.2 & -10.36 & 0.228 & 12.4 & 0.21 \\
        6 & 8 & 0 & 20 & 3788 & 50.3 & 0.326 & 0.233 & 0.209 & 0.108 & 0.039 & 0.031 &  & 5.8 & -13.6 & 0.173 & 10.7 & 0.37 \\
        6 & 9 & 0 & 28 & 8150 & 21.1 & 0.654 & 0.249 & 0.044 &  &  &  &  & 2.7 & -17.02 & 0.136 & 10.4 & 0.36 \\
        6 & 12 & 0 & 127 & 61108 & 57.5 & 0.255 & 0.243 & 0.173 & 0.104 & 0.09 & 0.026 & 0.023 & 8 & -28.4 & 0.079 & 10 & 0.34 \\
        7 & 4 & 0 & 3 & 126 & 38.5 & 0.463 & 0.364 & 0.173 &  &  &  &  & 2.8 & -4.22 & 0.645 & 13.4 & 0.48 \\
        7 & 8 & 0 & 31 & 8512 & 39.3 & 0.34 & 0.176 & 0.126 & 0.106 & 0.096 & 0.072 & 0.021 & 7.3 & -13.39 & 0.195 & 12.5 & 0.34 \\
        8 & 8 & 0 & 47 & 17575 & 24.1 & 0.446 & 0.153 & 0.127 & 0.086 & 0.074 & 0.028 &  & 6.3 & -14.46 & 0.184 & 12.9 & 0.36 \\
        9 & 8 & 0 & 71 & 33885 & 36.7 &  0.254& 0.154 & 0.113 &0.087  &0.067  & 0.056 &0.038  & 13.5 & -14.68 & 0.204 & 15.8 & 0.34 \\
 \hline
    \end{tabular}
    \caption{Properties and numerical results for various systems. For numerical results, 100,000 realizations of the two-body random ensemble were used. Columns 1,2, and 3 identify the system and the ground state spin of interest,
    Sec. \ref{sec:BE}. The next two columns show the number of states with spin $J$ in a given
    system, $D_{\ell N}(J)$, and the total number of spin states, $D_{\ell N},$ see Appendix  \ref{apx:B}. The following column shows $P(J)$ in percentage,
    see Sec \ref{sec:gs}. The following eight columns separated by double-lines correspond to the discussion in Sec. \ref{sec:gswfs}. 
    The  parameters $q_i$ listed in the
    descending order can be interpreted as the probabilities of a previously determined fixed $i$-th  state to be a ground state; $D_{\rm gs}$ is the effective
    dimensionality of space spanned by the ground state.
    The last four columns separated by the second double-line follow the discussion in Sec.~\ref{sec:energy}. They include two parameters of the Gumbel
    distribution in eq.~(\ref{eq:gumbel}) and, extracted from these parameters, the number of states competing to be in the ground state $\mathcal{D}$ and the width of the energy
    distribution $\sqrt{\overline{E^2}}/\sigma_{\rm max}$ relative to the  $\sigma_{\rm max}=N(N-1)/2.$
    \label{tab:main}}
\end{table}

\end{widetext}

%


\section{Ground-state wave functions\label{sec:gswfs}}

The previous analysis suggests that the states that appear as ground states are dominated by specific structures.
For the following, let us consider $W$ realizations from the random ensemble leading to a series of ground states
$|\varphi_n\rangle$ with a particular spin $J,$ where $n=1,\ 2,\dots ,\ W$ labels each individual realization.
As we are dealing with a particular angular momentum $J$, each of these wave functions can be expanded in $D=D_{\ell N}(J)$  basis states $|1\rangle$ which are eigenstates of $J^2$.
In order to explain our procedure, let us suppose that most wave functions in the series $|\varphi_n\rangle$ are nearly the same, being close to
some wave function $|\phi\rangle$ that can also be  expanded as
\begin{equation}
|\phi\rangle = \sum_1 c_1 |1\rangle.
\label{eq:mm}
\end{equation}
In order to find the best $|\phi\rangle$ from our ensemble,  we should maximize the sum of all squared overlaps as a function of the unknown set $\{c_1, c_2, \dots, c_D\}$
\begin{equation}
\frac{1}{W} \sum_n |\langle\varphi_n |\phi\rangle |^2 = \sum_{1, 2} c_1^{*} Q_{12} c_2 ,
\label{eq:mm1}
\end{equation}  
where the matrix element $Q_{12}$ is
\begin{equation}
 Q_{12}= \frac{1}{W} \sum_n 
\langle 1 | \varphi_n \rangle   \langle \varphi_n | 2 \rangle. 
\label{eq:mm2}
\end{equation} 
The solution is well known; the quadratic form in eq. (\ref{eq:mm1}) is maximized for the largest eigenvalue of matrix $Q$. 
The eigenvector corresponding to the largest eigenvalue provides the solution for the set of coefficients $\{c_1, c_2, \dots ,c_D\}.$

Let $q_i$ be a set of eigenvalues of $Q$ organized in descending order for  $i=1,2, \dots, D $ and $|\phi_i\rangle$ be the corresponding eigenvector.  
As follows from (\ref{eq:mm1}), the matrix is positive definite so all eigenvalues $q_i$ are positive. Moreover, 
as seen from eq.~(\ref{eq:mm2}), the trace is equal to one, so 
\begin{equation}
\sum_i q_i=1.
\label{eq:mm3}
\end{equation}
If all ground states $|\varphi_n\rangle$ actually had the same wave function, then the sum in eq.~\eqref{eq:mm1} would be equal to one and the maximal eigenvalue
of the factorized matrix \eqref{eq:mm2} would be $q_1=1$ while the remaining eigenvalues would be equal to zero. 

To highlight the meaning of these eigenvalues,  let us imagine that each ground state $|\varphi_n\rangle$ from the random ensemble always exactly coincides 
with one of the wave functions  $|\phi_i\rangle.$ 
Then, the eigenvalue $q_i$ represents a fraction describing how often a particular $i$-th state happens 
to be a ground state.

In general, this interpretation provides an assessment of the  dimensionality of space spanned by the ground states  $|\varphi_n\rangle.$ 
As seen in Table \ref{tab:main}, which includes the most prominent eigenvalues of matrix $Q$ for each system, 
in most situations there are only a few large eigenvalues while all other ones are small. This indicates that, while the Hilbert space dimensionality given 
by $D_{\ell N}(J)$ can be very large, only a small fraction of these states appear as ground states. 
The effective dimensionality of the space spanned by the ground states can be evaluated using entropy 
\begin{equation}
D_{\rm gs}=\exp(S),\, {\rm where}\quad  S=-\sum_i q_i \ln (q_i).
\end{equation}
The effective dimensionality $D_{\rm gs}$ is listed in Table \ref{tab:main} and is always much smaller than the total number of states of a given spin $D_{\ell N}(J).$

Let us carry out the analysis of the ground state wave functions to understand the minor preponderance of ground states with $J=(\ell-1)N.$ 
Consider the $\ell=5$ and $N=8$ system. As listed in Table \ref{tab:main}, the  $J=32$  ground state happens in about $1.1\%$ of realizations, which is a lot 
given only 7 states with this spin and 1514 possible spin states. 
Analysis of the eigenvalues of matrix $Q$ shows that only two are effectively non-zero.  It turns out that in this case the 
ground state wave function is almost exclusively one of two possibilities, with corresponding probability for $|\phi_1\rangle$ being  51\% and the probability for $|\phi_2\rangle$ being 49\%. The realizations with $|\phi_2\rangle$ have $|\phi_2\rangle$ as a ground state wave function exactly, 
with no admixtures.   Some very small admixtures are present in realizations with $|\phi_1\rangle.$ Assuming linearity and following 
Eqs. \eqref{eq:23}, \eqref{eq:25}, and  \eqref{eq:26} we find $\sqrt{\overline{E^2}}/\sigma_{\rm max}= 0.77$ and 0.71 for $\phi_1$ and $\phi_2,$ 
respectively. The $\phi_1$ relies on attractive matrix element $V_{10},$ while $\phi_2$ emerges in the ground state due to attraction in $V_{8}.$

\begin{figure}[h]
\begin{center}
\includegraphics[width=1.0\linewidth]{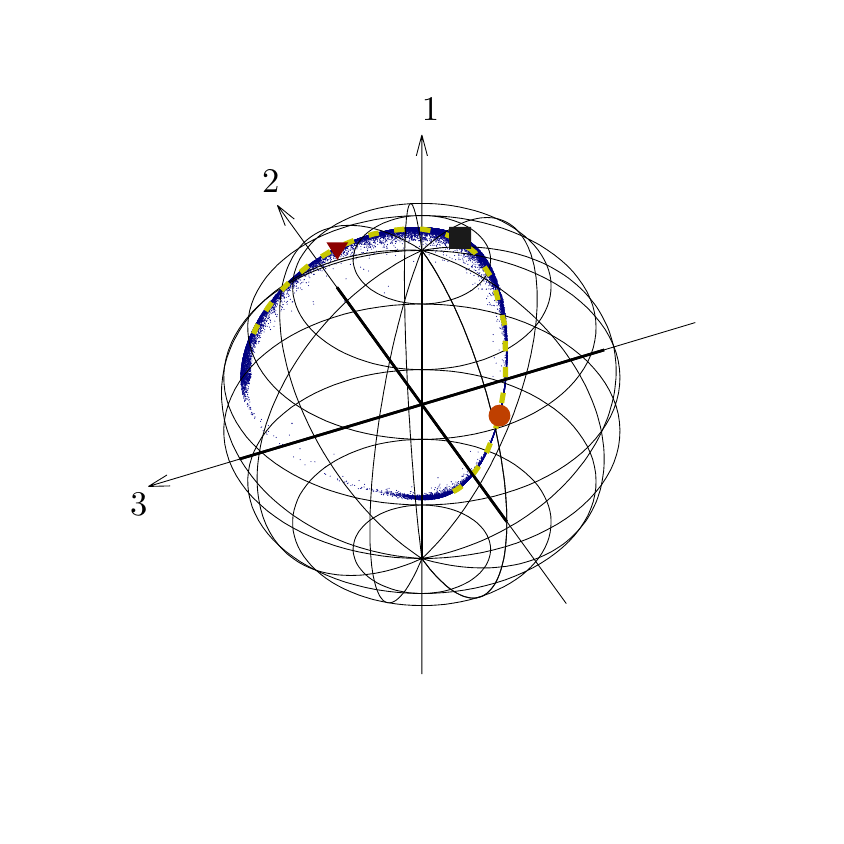}
\end{center}
\caption{Ground state wave functions with $J=0$ for a system of $N=8$ $\ell=5$ bosons are shown for the two-body random ensemble on a 3-dimensional sphere. The dashed line traces the line formed by the  $J=0$  wave functions of the Hamiltonians containing  only $V_0$ and $V_2$ matrix elements. 
Ground states corresponding to the quadrupole-quadrupole interaction Hamiltonian (square),  Hamiltonian where $V_2=-1$ while 
everything else is zero (triangle), and pairing (circle) are shown. 
\label{fig:sphere}  
}
\end{figure}

For the same $\ell=5$ and $N=8$ system, the $J=0$ ground
state happens in nearly $34\%$ of random realizations. There are 12 $J=0$ states in the system, and yet the effective dimensionality is only 
$D_{\rm gs}=3.$ This low dimensionality allows us to visualize the wave functions $|\varphi_n\rangle$ in 
Figure \ref{fig:sphere} using a three-dimensional unit sphere. 
The $n$-th wave function is shown by a point defined by the three components  $\langle \phi_1 |\varphi_n\rangle,$ $\langle \phi_2 |\varphi_n\rangle,$ 
and $\langle \phi_3 |\varphi_n\rangle$, which are the overlaps of $|\varphi_n\rangle$ with the principal eigenvectors of
the $Q$ matrix. The phase is selected so that the first component is positive, thus all points are on the upper hemisphere. 
It is remarkable that the points are not covering the hemisphere uniformly; rather, they form a curve on the sphere, which indicates even further reduction
of the measure of space spanned by the ground states. Considering $J=0$ ground states of the Hamiltonians where only $V_0$ and $V_2$ are non-zero 
allows one to trace this curve, shown by the dashed line in 
Figure \ref{fig:sphere}. Special cases corresponding to quadrupole-quadrupole interaction (square), pairing (circle), and $V_2=-1$ while 
everything else is zero (triangle) are shown.  The highest density of points is in the vicinity of the quadrupole-quadrupole Hamiltonian ground state 
(square), while around pairing (circle) and $V_2=-1$ (triangle) the density is low.

While some of these findings are specific to the systems considered, the analysis of ground state wave functions shown in this section highlights that, 
out of the entire space of wave functions of a given $J$, those that appear as ground states span only a small subspace. Their structures are generally 
determined by a few matrix elements and their energy distribution variances, studied in the previous section, suggest condensate type structures.

\section{Clustering}

As shown above, the ground states of randomly interacting systems of bosons are not uniformly random vectors in the Hilbert space; they are special, 
condensate-type structures and these properties allow them to have lower energy.  The example of $d$-bosons in Sec. \ref{sec:dbos}, where all $J=0$ 
states are either pair or triplet condensates, suggests that cluster condensates may be a general feature of boson systems. Oscillatory behavior as a function of the particle 
number seen in  Fig. \ref{fig:4} also supports the idea of triplets, quartets, or perhaps even bigger structures playing a role.  

In what follows we limit our study to $J=0$ ground states in systems of $\ell=6$ bosons. 
If, similar to $d$-bosons, the $J=0$ ground states are condensates, then systems with different 
numbers of clusters (but with the same Hamiltonian) should be similar. 

\begin{figure}[h]
\begin{center}
\includegraphics[width=0.9\linewidth]{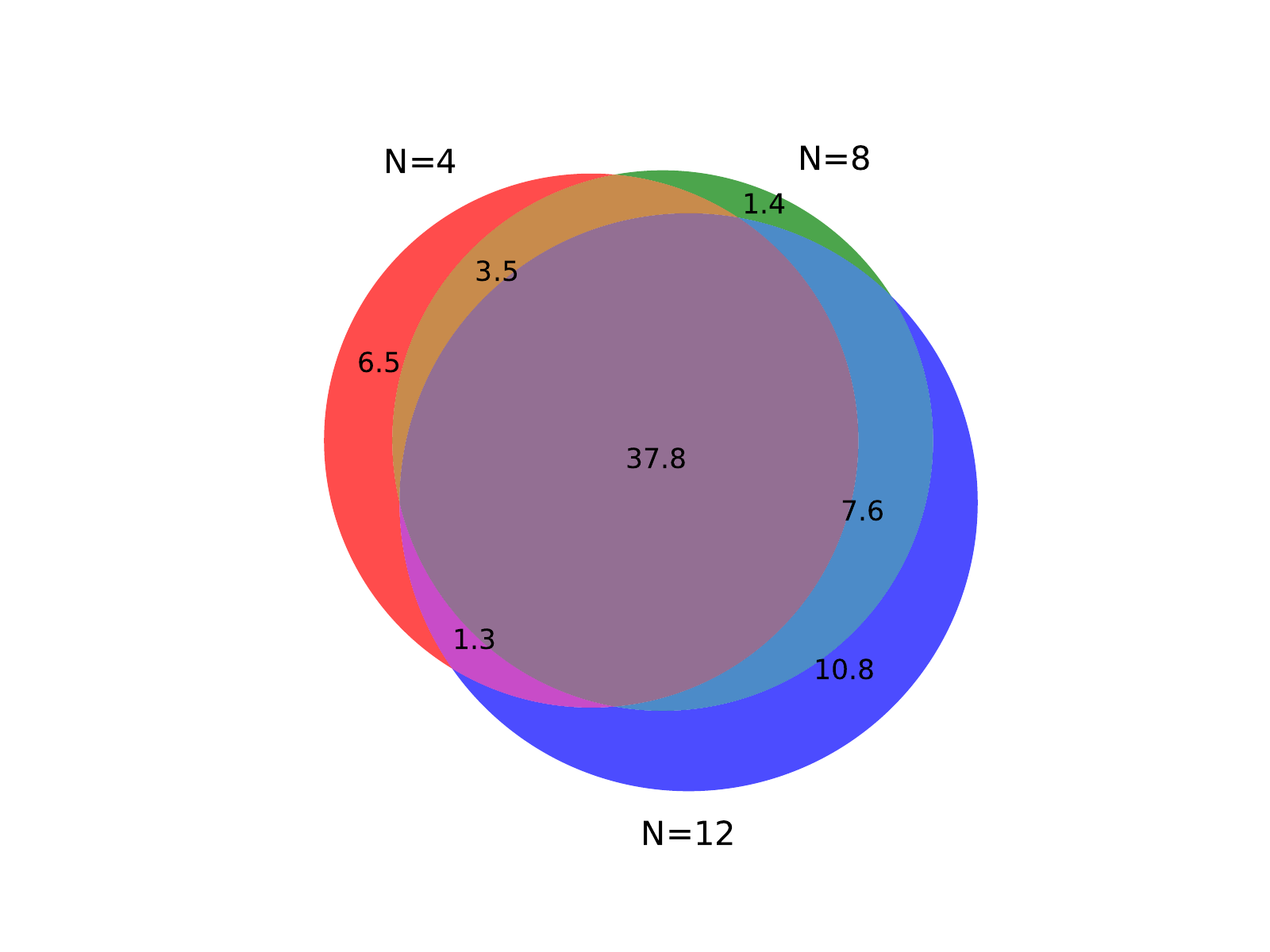}
\end{center}
\caption{Venn diagram showing the relationship between the sets of Hamiltonians from the two-body random ensemble that have a $J=0$ ground state
in the $N=4$, $N=8$, and $N=12$ systems of $\ell=6$ bosons. The numbers show percentage of the corresponding Hamiltonians from the total 
ensemble. The diagram is drawn to properly reflect the scale of sets and their overlaps.  
\label{fig:venn}  
}
\end{figure}

Let us review the sets of Hamiltonians and their overlaps that produce the $J=0$ ground states for various particle numbers.
In Figure \ref{fig:venn}, using a Venn diagram (also known as a set diagram),
we show the set of Hamiltonians comprising the two-body random ensemble that amount to $J=0$ ground states
in $N=4,$ $N=8,$ and $N=12$ systems of $\ell=6$ bosons. The diagram reflects the sizes of the sets and their overlaps. All three
systems are very similar.  For example, the cases where, with the same interaction Hamiltonian, an $N=4$ system has a $J=0$
ground state but the ground state spin of the $N=8$ system is non-zero, are rare.

\begin{figure}[h]
\begin{center}
\includegraphics[width=0.9\linewidth]{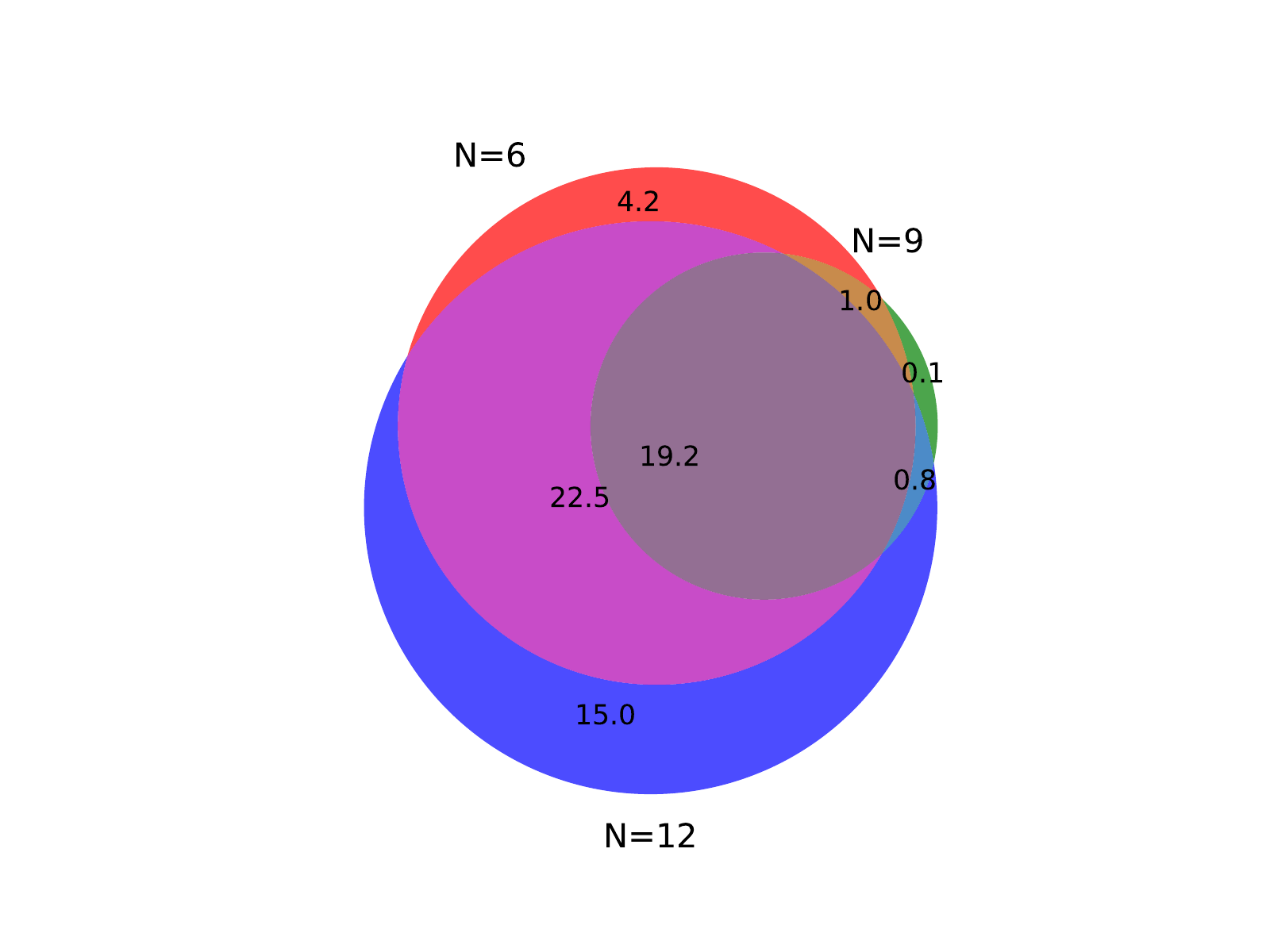}
\end{center}
\caption{Venn diagram, similar to Fig. \ref{fig:venn},  showing the set of Hamiltonians from the two-body random ensemble that have a $J=0$ ground state
in the $N=6$, $N=9$, and $N=12$ systems of $\ell=6$ bosons. The numbers show percentage of the corresponding Hamiltonians from the total 
ensemble. The diagram is drawn to properly reflect the scale of sets and their overlaps.  
\label{fig:venn3}  
}
\end{figure}

A slightly different situation is seen in Figure \ref{fig:venn3} that compares the sets for $J=0$ states in the $N=6,$ $N=9,$ and $N=12$ 
systems of $\ell=6$ bosons.  All sets overlap covering $N=9.$ The $N=3$ set is not shown in this figure, but its overlap with $N=9$ 
is at the 90\% level, indicating that in 
about 17\% of overall cases we are dealing with spin zero triplets.
 
%

\subsection{Pairing}
It appears that pairs, triplets, and  quartets are the most likely types of clusters. It is most instructive to start with pairs and pairing. 
Given that paired ground states are common in realistic situations, such as the superconducting ground state of fermions \cite{broglia:2013} or as a solution of the 
interacting  boson model for nuclei \cite{arima:1981a}, the prevalence of paired states is natural to expect. However, numerous studies with randomly interacting fermions 
show this not to be the case \cite{zelevinsky:2006:art,zelevinsky:2001:art,mulhall:2000:art}. Our studies also show pairing not to be prevalent  for randomly interacting bosons,  with the exception of some very restrictive situations, such as with  $d$-bosons.  
As was already mentioned in the discussion related to Fig. \ref{fig:sphere}, the number of realizations with paired ground states is low. This is evident from the 
low density of points around the paired state (orange circle) in Fig. \ref{fig:sphere}.   

\begin{figure}[h]
\begin{center}
\includegraphics[width=0.9\linewidth]{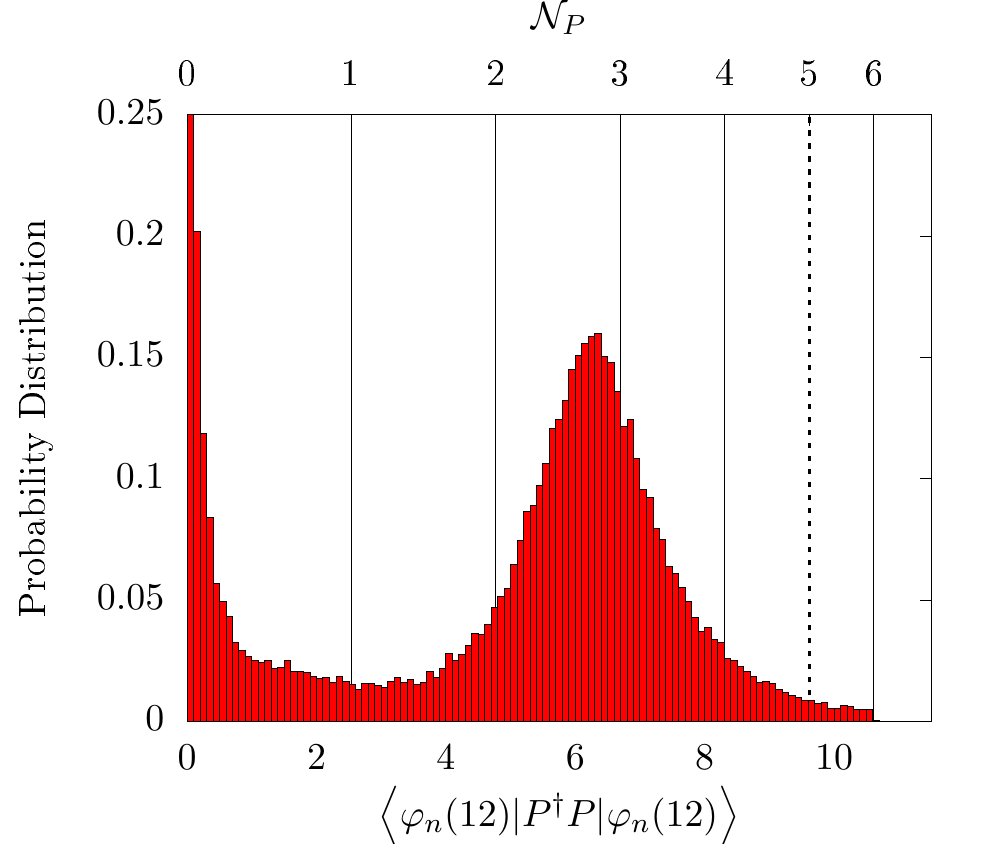}
\end{center}
\caption{The distribution of the pair number is shown for  the $N=12$ system of $\ell=6$ bosons. The vertical lines 
show the values from eq. \eqref{eq:19}  for $\mathcal{N}_P=0,1,2,\dots, 6$ pairs. Technically, $\mathcal{N}_P=5$ is not allowed for $J=0$ states but its location is shown for reference. 
\label{fig:npair}  
}
\end{figure}

In Fig.~\ref{fig:npair}  
we show the probability distribution for  $\langle\varphi_n(N)|P^\dagger P|\varphi_n(N) \rangle$ (where we denote $P \equiv P_{00}$) for the $N=12$ particle system.
Since the eigenvalues of the pairing Hamiltonian are known analytically, eq. \eqref{eq:19}, and are associated via the seniority $\nu$ with the number of pairs,
${\cal N}_P=(N-\nu)/2$ we  can assess the typical number of pairs. As evident from Fig.~\ref{fig:npair}, 
there are many cases 
(about a quarter of realizations when the ground state is $J=0$)
with no pairs at all,  
while most of the remaining systems have about 3 pairs on average. This is far from the maximum of 6 pairs that corresponds to   
$\langle P^\dagger P\rangle=138/13\approx 10.6.$

\subsection{Triplets\label{sec:TripSec}}

In a system with even $\ell$, the
three-boson state with spin $J=0$ is unique, as mentioned in Appendix \ref{apx:A2}. This allows us to uniquely define a triplet creation
operator $T^\dagger\equiv T^{\dagger}_{00}$ that  creates this state from the vacuum. Using the triplet creation and annihilation operators, we can assess the level of triplet clustering
in the ground states and whether the ground states of different particle number are connected by triplet removal and addition. 

\begin{figure}[h]
\begin{center}
\includegraphics[width=0.9\linewidth]{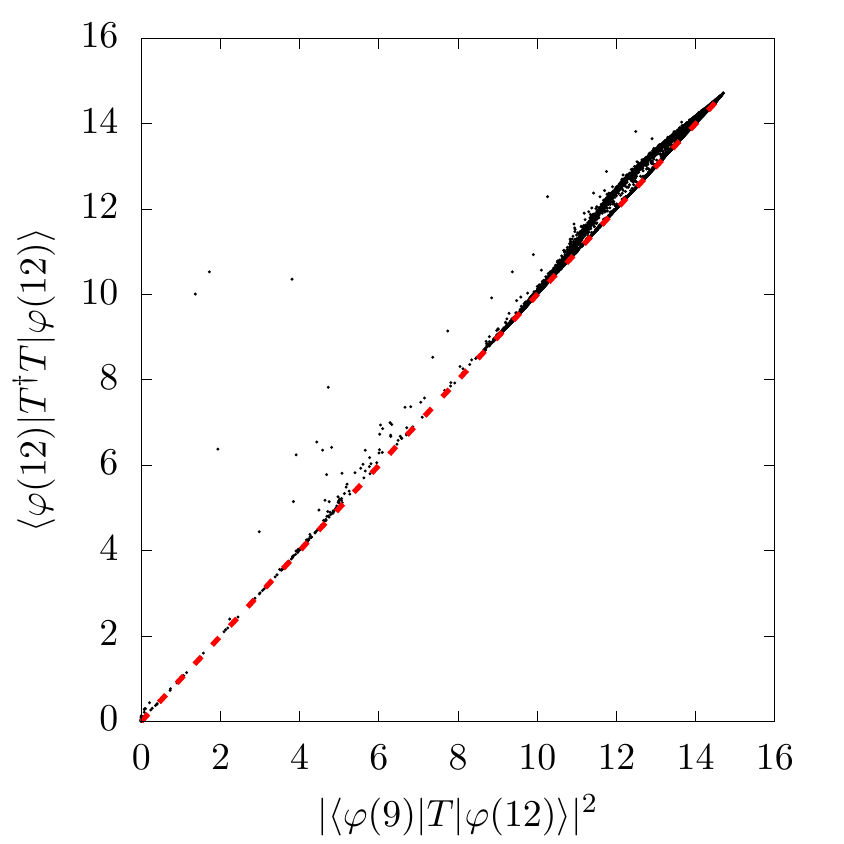}
\end{center}
\caption{Scatter plot that, for a 12-particle system of $\ell=6$  bosons, shows triplet removal amplitude squared 
$|\langle\varphi(9)|T|\varphi(12) \rangle|^2$ versus normalization
$\langle\varphi(12)|T^\dagger T|\varphi(12) \rangle.$ When triplet removal identically leads to the ground state of the $N=9$ system 
$|\varphi(9)\rangle = T |\varphi(12) \rangle$, the two quantities are equal; this condition is shown by a dashed diagonal line. 
\label{fig:tm}  
}
\end{figure}

In Fig.~\ref{fig:tm} we examine the removal of a triplet  from the $J=0$ ground state of the $N=12$ particle system $T|\varphi_n(12) \rangle.$ For each $n$-th realization 
of the ensemble, using a scatter plot, we show  the overlap of the resulting state with the ground state of the $N=9$ particle system squared
$|\langle\varphi_n(9)|T|\varphi_n(12) \rangle|^2$ on the $x$ axis and  $\langle\varphi_n(12)|T^\dagger T|\varphi_n(12) \rangle$ on the $y$
axis; the latter represents the norm of the state after triplet removal.
All points in the figure appear very close to the diagonal line, meaning that the two quantities are nearly equal; thus removal of a  triplet
from the $N=12$ system leads to the ground state of the $N=9$ system.  Using a complete set of eigenstates  in the $N=9$ particle system
labeled by $i$, the norm on the $y$ axis can be expanded as 
\begin{equation}
\langle\varphi_n(12)|T^\dagger T|\varphi_n(12) \rangle=\sum_i |\langle\varphi^{(i)}_n(9)|T|\varphi_n(12) \rangle|^2.
\end{equation}
The result shows that a single term with the index $i$ corresponding to the ground state dominates this sum. 

\begin{figure}[h]
\begin{center}
\includegraphics[width=0.9\linewidth]{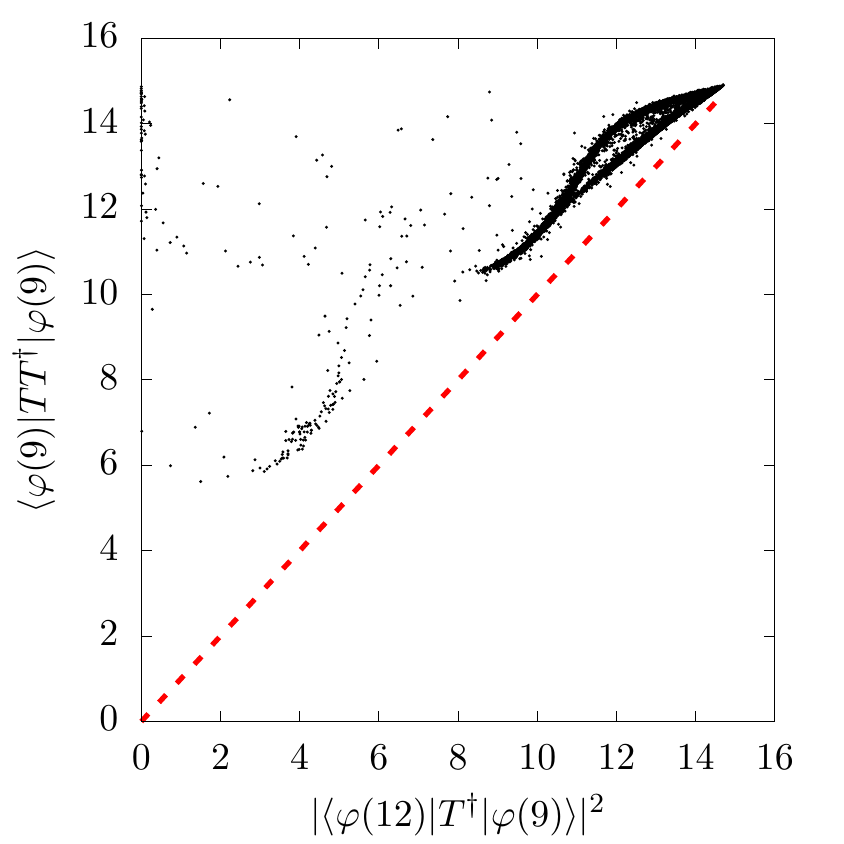}
\end{center}
\caption{This figure is similar to Fig.~\ref{fig:tm} but considers  triplet addition to the $N=9$ system; 
$|\langle\varphi(12)|T^\dagger|\varphi(9) \rangle|^2$ versus normalization
$\langle\varphi(9)|T T^\dagger|\varphi(9) \rangle.$
When triplet addition leads to the ground state of the $N=12$ system 
$|\varphi(12)\rangle = T^\dagger |\varphi(9) \rangle$, the two quantities are equal and the scattered points appear on a  diagonal line. 
\label{fig:tp}  
}
\end{figure}

In  Figure~\ref{fig:tp}  we show the addition of a triplet to the 9-particle system, $T^\dagger|\varphi_n(9) \rangle.$
The results are similar, although some deviations indicate that the ground states of 12 particle systems have some small additional components that
make $|\varphi_n(12)\rangle$ slightly different from $T^\dagger|\varphi_n(9) \rangle.$ More deviations in the case of cluster addition as compared to cluster 
removal appear to be a general feature that we observed for other systems as well. They could be caused by the presence of different cluster types, 
we discuss quartets in what follows, or by other phenomena. 

\begin{figure}[h]
\begin{center}
\includegraphics[width=0.9\linewidth]{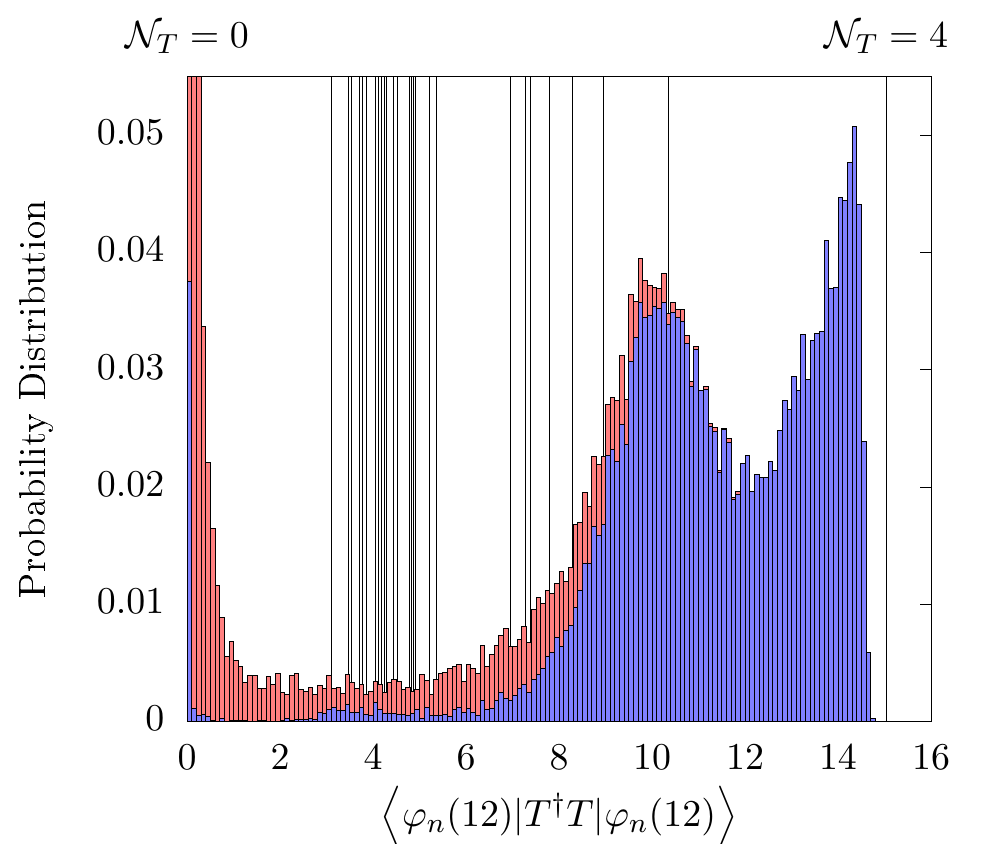}
\end{center}
\caption{The distribution of the triplet number is shown for the $N=12$ system of $\ell=6$ bosons. The vertical lines 
show eigenvalues of the triplet number operator.  The histogram in red shows all { 58}\% of realizations with a $J=0$ ground state;  the histogram in blue corresponds to those 19\% of ensemble realizations  where $N=6$ and 9 particle systems also have a $J=0$ ground state. 
\label{fig:number}  
}
\end{figure}

In order to study the number of  triplets in the ground states, in Fig.~\ref{fig:number} we show the distribution of 
$\langle\varphi_n(12)|T^\dagger T|\varphi_n(12) \rangle$, which appeared on the y-axis in Figure \ref{fig:tm} for the $N=12$ system.
Similar to pairing studied in Fig. \ref{fig:npair}, this quantity is to be interpreted as the cluster number. However, unlike for pairing, there are 
no analytic eigenvalues for the triplet number operator $T^\dagger T$.  Therefore, we diagonalize $T^\dagger T$ numerically. 
Similar to Figure \ref{fig:npair}, the eigenvalues of $T^\dagger T$ are shown by vertical lines in Fig.~\ref{fig:number}.
We expect the $J=0$ states in the 12 particle system to have 0,1,2 or 4 triplets.  Note that a spin zero state with three triplets is not possible
because three remaining particles must also have spin zero and thus form a fourth triplet.
Out of 127 $J=0$ states in the system, there are 100 zero eigenvalues that we associate with no triplets at all, ${\cal N}_T=0$.
The unique largest eigenvalue of about
15 clearly corresponds to the full condensate of ${\cal N}_T=4$ triplets.
The remaining eigenvalues correspond to intermediate situations but, given clear visible gaps,
we can roughly assign those between about 7 and 10 as corresponding to two triplets and those between about 3 and 5 to one triplet.
The red histogram shows all cases (about 60\% of realizations) when $J=0$ is the ground state
in the 12-particle system.
Out of these, the blue histogram shows the cases when the same realization also gives a $J=0$ ground state in the $N=6$ and $N=9$ systems,
which is  about 20\% of realizations, see Venn diagram \ref{fig:venn}. 
The peak near the maximum triplet number indicates that many of these systems (about 10\% of all realizations) are nearly a perfect triplet condensate state. 
Note that this result is very different from that seen in Fig. \ref{fig:npair} for pairing. 

The other peak in the probability 
distribution near zero for red cases shows that, where most realizations of 6 or 9 particle systems or both do not have a spin $J=0$ ground state, the 
$J=0$ ground state of the 12 particle system has no triplets. The fraction of these realizations that possess $J=0$ ground states along with $N=4$ and $N=8$ systems, see Fig.~\ref{fig:venn}, is about 38\% and the peak in Fig.~\ref{fig:number}
near zero ($\langle \varphi_n(12)|T^\dagger T|\varphi_n(12)\rangle<0.8 $ ) comprises 35\% of realizations. 
As we discuss in the following subsection, the 
structure of those systems is dominated by quartets.

\subsection{Quartets}

\begin{figure}[h]
\begin{center}
\includegraphics[width=0.9\linewidth]{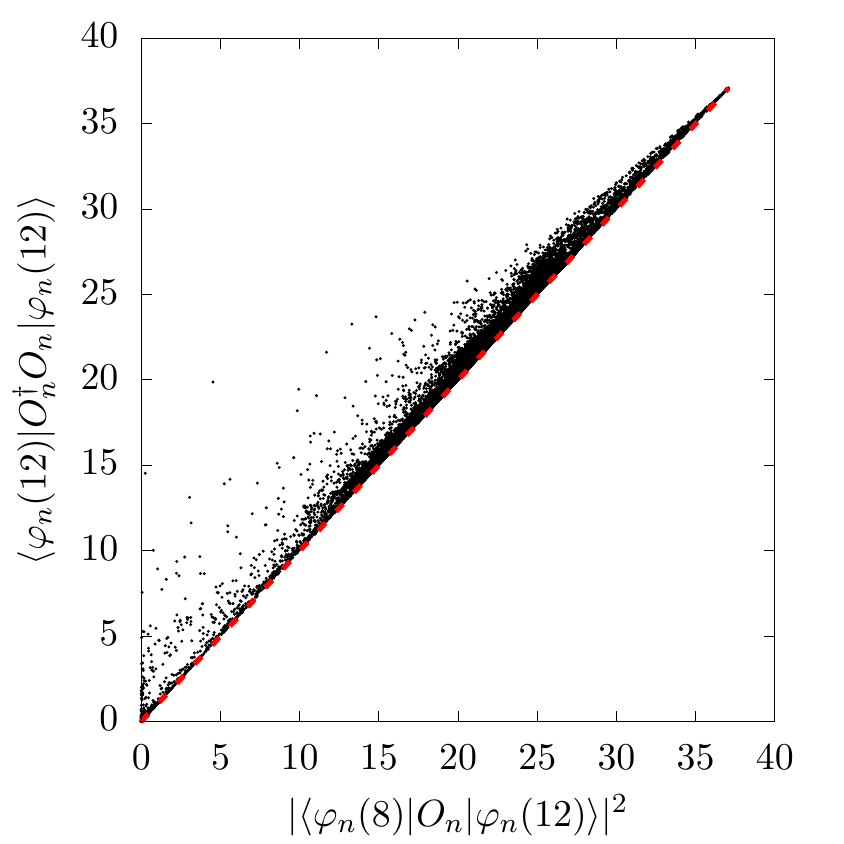}
\end{center}
\caption{This figure is similar to Fig.~\ref{fig:tm} but considers a quartet removal from an $N=12$ system. Unlike with the triplets, the quartet is defined 
for each random realization using a ground state of a 4-particle system. Only those cases are studied where all systems involved, the $N=4, 8$ and 12 systems, have 
a $J=0$ ground state. This happens in 37.8\% of realizations, as seen in the Venn diagram  Fig.~\ref{fig:venn}.
\label{fig:qm}  
}
\end{figure}

\begin{figure}[h]
\begin{center}
\includegraphics[width=0.9\linewidth]{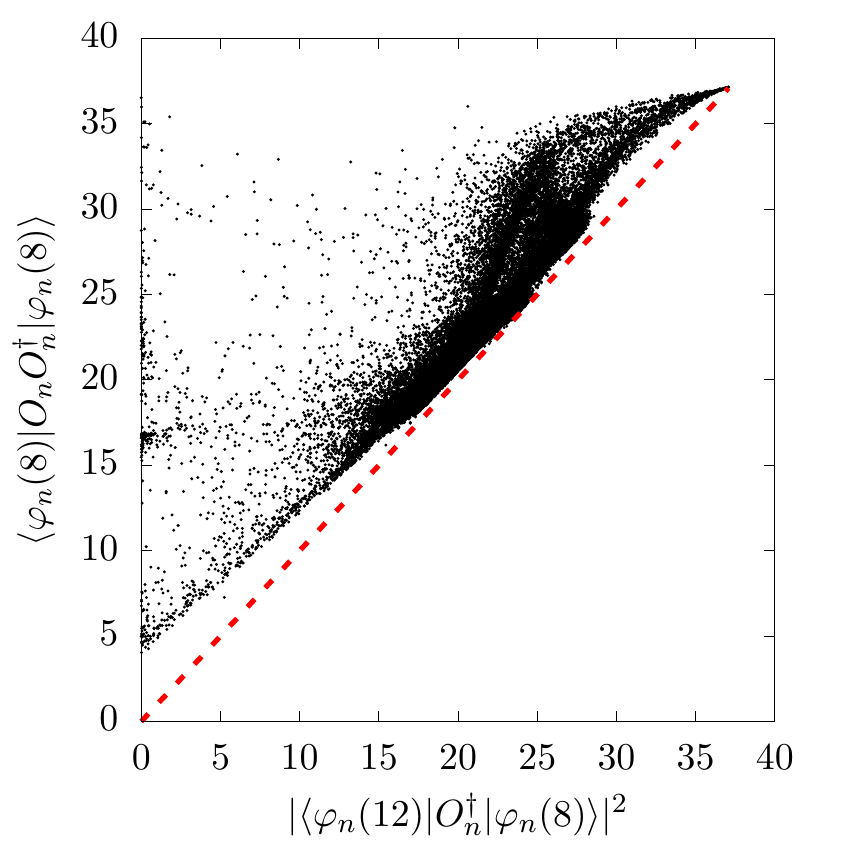}
\end{center}
\caption{This figure is similar to Fig.~\ref{fig:qm}, but for quartet addition to the ground state of the $N=8$ system. 
\label{fig:qp}  
}
\end{figure}

Quartets are more difficult to address since, generally, an $N=4$ system has several states with spin $J=0,$ for example, $D_{\ell=6,N=4}(0)=3$.
One of these states is associated with pairing while the other two offer different types of possible quartets. Whichever structure dominates in the ground state
is determined by the Hamiltonian. As seen from the data in Tab. \ref{tab:main}, the effective dimensionality $D_{\rm gs}= 2.8$,
so all three $J=0$ states appear as ground states relatively often.
Thus, for the study of quartets we define the quartet operator individually for each realization
using the corresponding $J=0$ ground state of the four-particle system, so that
\begin{equation}
O^\dagger_n|0\rangle\equiv |\varphi_n(4)\rangle.
\end{equation}
Our results in Figs. \ref{fig:qm} and \ref{fig:qp}, similar to the previously considered  triplets in Figs. \ref{fig:tm} and \ref{fig:tp}, show that the ground states of these systems are close to those formed by a repeated action of the quartet operator: 
\begin{eqnarray}
|\varphi_n(8)\rangle \propto O^\dagger_n |\varphi_n(4)\rangle = \left (O^\dagger_n\right)^2 |0\rangle, \\ 
|\varphi_n(12)\rangle \propto O^\dagger_n |\varphi_n(8)\rangle \propto \left (O^\dagger_n\right)^3 |0\rangle.
\end{eqnarray}
Thus, the ground states in larger systems are created from an $N=4$ ground state by replicating the quartet several times, which of course is not a problem for bosons. 

Finally, commenting on the structure of the $N=12$ systems, we can summarize that the $J=0$ ground state happens in about $58\%$ of cases, and out of those
16\% are dominated by triplet structure (with 10\% being in nearly perfect four-triplet condensate state) and about 38\% by quartets. The overlap between the two types of structures is small. 

\section{Conclusions}
Random two-body ensembles provide a unique perspective on the general emergence of phenomena in quantum many-body physics where out of randomness
and complexity unique features, new degrees of freedom, and collective dynamics emerge. 
The main finding of this work is that the ground states of bosonic two-body random ensembles are not actually random. 

First, what has been known for 
some time, the statistics of spins of states that appear as ground states are dominated by spin zero for even-particle systems and by a single-boson 
spin for odd-particle systems. The chances of seeing the most aligned state with the maximum spin possible are also enhanced, but the coherent boson-condensate structure of that is known. 

Second, our numerical study of the ground state energy distribution using extreme value distribution theory shows that out of all states in the
spectrum only about a dozen actually compete to be in the ground state. Moreover, all of these states are collective and their energies scale with the number of pairs
$\propto N^2.$

Third, the analysis of the ground state wave functions shows that they span a very small subspace out of the total 
allowed Hilbert space.
As seen in Table \ref{tab:main}, roughly 80\% of ground state wave functions are comprised of linear combinations of two components. 

Finally, we find that the above results for $J=0$ are largely explained by the formation of condensates of clusters, mainly dominated by 
spin-zero triplets and quartets. 
For spin-zero triplets, which are uniquely defined, numerical studies show that ground states of adjacent systems 
$N$ and $N\pm3$ are connected by triplet removal 
and triplet addition; and 
study of the distribution of the triplet number operator support that a large fraction of ground states are triplet condensates. 
Similar results are seen for quartets; in that case the $J=0$ ground states of large systems with $N=8$ and 12  have a structure 
built from the $N=4$ system by repeating it 2 and 3 times, respectively. 

There are certainly questions that remain outside the scope of this study, including how these results extrapolate for larger systems and whether larger clusters play a significant role. 
It is also interesting that while the interaction is two-body, pairing does not play a more significant role; reorganization of particles into clusters of more than two particles clearly appears to be more favorable. 
The main conclusion of our study is that the emergence of correlated structures in quantum many-body systems of identical particles is highly probable  even though their interactions are random.
 
\begin{acknowledgments}
This material is based upon work supported by the U.S. Department of Energy Office of Science , Office of Nuclear Physics under Grant No. DE-SC0009883
\end{acknowledgments}

\appendix
\section{Bosonic geometry\label{apx:A}}

The question of spin statistics, namely, how many states of a given spin there are in a many-body 
system
addresses the ``geometry'' of the Hilbert space, which is independent from the interaction Hamiltonian. 
We define $D_{\ell N}(J)$ to be the number of states with spin $J$ (multiplets of $2J+1$ projections) in the system of $N$ identical bosons of spin $\ell.$   
Then, the dimensionality of the Hilbert space is 
\begin{equation}
D'=\sum_J (2J+1) D_{\ell N}(J)= \frac{(2\ell+N)!}{(2\ell)! N!},
\label{eq:2}
\end{equation}
where the prime indicates a quantity that accounts for magnetic substates.
For convenience of comparison we define the total number of states in the system (not including 
magnetic degeneracies) as 
\begin{equation}
D_{\ell N}=\sum_J D_{\ell N}(J),
\label{eq:3}
\end{equation}
and the fraction of states of a given $J$ as
\begin{equation}
d(J)= \frac{D_{\ell N}(J)}{D_{\ell N}}.
\label{eq:4}
\end{equation}

\begin{figure}[h!]
\begin{center}
\includegraphics[width=0.8\linewidth]{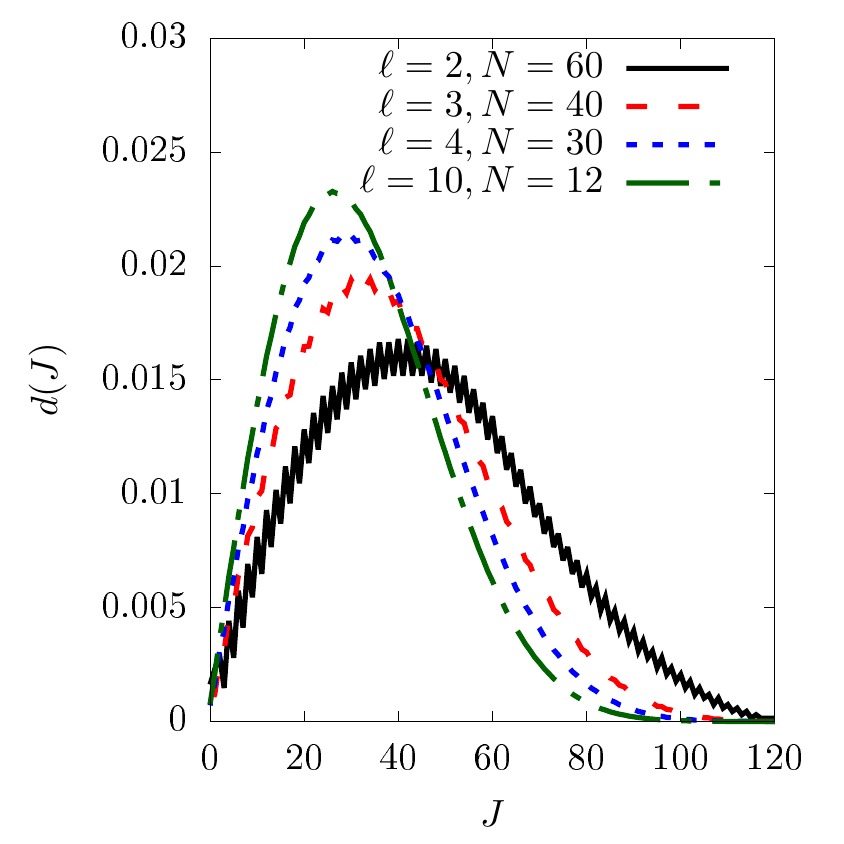}
\end{center}
\caption{\label{fig:disbos} The fraction of states of a given spin $J$ is shown for several different systems.
}
\end{figure}

In Fig.~\ref{fig:disbos}, we show $d(J)$ as a function of $J$ for several systems. This is a peaked distribution showing that the fraction of states with low spins and high spins is low.  

The distribution of magnetic projections $M$, which is the sum of magnetic projections of individual particles, due to the central limit theorem, is expected to be nearly Gaussian, see Refs. \cite{ericson:1960,zelevinsky:2017}. Then, having 
\begin{equation}
D'(M)=D'\sqrt{\frac{\beta}{\pi}} e^{-\beta M^2},
\label{eq:5}
\end{equation}
the number of states with a given $J$ is 
\begin{equation}
D_{\ell N}(J)=D'(M=J)-D'(M=J+1),
\label{eq:6}
\end{equation} 
which results in
\begin{equation}
D_{\ell N}(J)\approx D'  \beta \sqrt{\frac{\beta}{\pi}}(2J+1)\, e^{-\beta J(J+1)}
\label{eq:7}
\end{equation} 
and 
\begin{equation}
d(J)\approx \beta (2J+1)\, e^{-\beta J(J+1)}.
\label{eq:7x}
\end{equation} 
This describes well the results shown in Fig.~\ref{fig:disbos}.

The bosonic nature of particles allows only for fully symmetric states. This restriction amounts to slight modification of the 
distribution (\ref{eq:5}), making it sub-Gaussian (platykurtic) with reduced tails. 


\subsection{Special structures}

The fully aligned state with the maximum spin $J_{\rm max}$ is unique 
\begin{equation}
D_{\ell N}(J_{\rm max})=1,\,\,{\rm where}\quad J_{\rm max}=N\ell.
\label{eq:9}
\end{equation}
For any unique state when $D_{\ell N}(J)=1$, the energy is a linear function of the interaction parameters $V_L.$
For an aligned state, a condensate of  bosons all with the same maximum or minimum magnetic projections, the energy from two body interactions is
\begin{equation}
E=\frac{N(N-1)}{2} V_{2\ell}.
\label{eq:10}
\end{equation}
The factor $N(N-1)/2$ reflects the number of pairs in the condensate of aligned bosons. 

There is no state with $J=\ell N-1$ and
there are always unique states with $J=\ell N-2$ and with $J=\ell N-3.$

A remarkable symmetry exists between systems of even number $N$ bosons with spin $\ell$ and $2\ell$ bosons each 
with spin $N/2$, namely between  $(\ell, N)$ and $(\ell'=N/2, N'=2\ell).$ 
As follows from eq.~(\ref{eq:2}), these systems have an identical number of states and an identical 
maximum $J_{\rm max}=\ell N=\ell' N'.$ These two Hilbert spaces break down into an identical number of irreducible representations of the rotational group. Thus, 
\begin{equation}
D_{\ell,\, N}(J)=D_{N/2,\, 2\ell}(J), {\rm where}\,\, N\,\, {\rm is\,\,even}.
\label{eq:11}
\end{equation}

In general, the number of states with a certain angular momentum is not known analytically, but there are special cases. For $N=2$, the result is trivially $D_{\ell N=2}(J)=1$ and  $D_{\ell N=2}=\ell+1.$

\subsection{System of three bosons \label{apx:A2}}
For three bosons, the sequence $D_{\ell, N=3}(J)$ as a function of $J$ is actually universal 
up to $J=\ell+1$, which is to say that 
 $D_{\ell, N=3}(J)$ does not depend on $\ell$ as long as $J\le \ell+1.$  We found a simple relation, 
\begin{align}
 \label{eq:12}
 D_{\ell, N=3}(J) &= 
 D_{\ell-2, N=3}(J)+ 
 \\
& \left \{ 
 \begin{array}{lc}
 1& {\rm for}\,\,\ell\le J \le 3\ell\,\, {\rm and} \,\,J\ne3\ell-1 \cr
 0& {\rm otherwise}
 \end{array}
 \right. ,
 \nonumber
 \end{align}
 that allows one to establish the number of states of each spin analytically. 
The above relation also implies that the sequence is unique near the terminating highest angular momentum $J=3\ell;$
in particular, $D_{\ell, N=3}=1,0,1,1,1,1,2...$ for $J=3\ell,\  3\ell-1,\ 3\ell -2, \dots \ .$

For $N=3$ odd-parity bosons, the recurrence (\ref{eq:12}) starts with $\ell=1$ implying that three  bosons 
of any odd angular momentum $\ell$
cannot couple to  $J=0$ and to $J=2$, which is similar to Furry's theorem.

For three bosons of even parity (starting with $\ell=2$), $J=1$ is not possible and states with $J=0,2,3$ and 5 (assuming $\ell>2$) are unique.  The energy of the unique $J=0$ state is 
\begin{equation}
E_0=3V_\ell.
\label{eq:13a}
\end{equation}

\subsection{Spin $\ell=1$ bosons}
For a system of bosons with $\ell=1$, the $D_{\ell=1 N}(J)=1$ for
$J=N,\ N-2, \dots, \ 0\,\ {\rm or}\,\ 1 $ and zero otherwise.
This property is well known for the case of the three-dimensional harmonic oscillator where, for each shell,
$N=2n+J$, where the number of quanta $N$ is represented by the number of bosons and $n$ is an integer.
It is also possible to think about the structure of aligned states with magnetic projection $M=J$ as being a two-condensate system: $n$ boson pairs coupled to $L=0$  are combined with an aligned state of $N-2n$ spin $\ell=1$
bosons, giving a total angular momentum of $J=N-2n.$

\subsection{Spin $\ell=2$ bosons}

The number of different spin  states for $d$ bosons can be worked out by considering it as a mixture of two
spinless condensates with free particles:
${\cal N}_P$ of $J=0$ pairs and ${\cal N}_T$ of $J=0$ triplets so that
$N=2{\cal N}_P+3{\cal N}_T+f,$ where $f$ represents the number of remaining particles not included into the spinless condensates.  These uncoupled $f$ particles are the ones producing the angular momentum
\begin{equation}
J=2f,\ 2f-2,\ 2f-3,\ 2f-4,\dots,\ f
\label{eq:14}
\end{equation}
which can take all integer values between $2f$ and $f$ with the exception of $2f-1.$

It is clear that similar condensates of spinless clusters appear as a generic feature of bosonic many-body states, but the number of different kinds of spinless clusters grows very fast and, at some moment, the condensates are no-longer orthogonal.
For $\ell=2$, the number of $J=0$ states can be found as the number of ways the total number of particles can be broken into
pairs and triplets, i.e. the number of all possible pairs $\{{\cal N}_P,{\cal N}_T\}$ so that $2{\cal N}_P+3 {\cal N}_T=N.$

\subsection{Spin $\ell=3$ bosons}
For $\ell=3$, there is no known analytical result, but organization into clusters is still useful.  
Considering $D_{\ell=3 N}(0)$ as a function of the particle number $N$, we find that, up to  
$N=30$ particles, all states with $J=0$  can 
be represented by spin-less clusters with sizes of 2, 4, 6, 10, and 15 bosons. Only  for $N=30$ $D_{\ell=3,N=30}(0)=47$ where as 
there are 48 different sets of 
$\{{\cal N}_2,{\cal N}_4,{\cal N}_6,{\cal N}_{10}, {\cal N}_{15}\}$ possible. Here we use ${\cal N}_{n}$ to denote a number of 
spineless clusters of $n$ bosons (${\cal N}_{2}={\cal N}_{P}$ in our other notation). 
Most likely, this implies that a pair of two spin-less 15-boson clusters can be 
represented as condensates of other types. 

For $\ell=3$, the largest system with no $J=0$ states is for $N=13,$ $D_{\ell=3\, N=13}(0)=0.$

\subsection{Other special cases} 
For $N=4$,  the states with $J=4\ell-4$ and $J=4\ell-5$ are unique.
Other nontrivial cases worth mentioning are $(\ell, N)=(5,5), (5,7)$ and $(7,5),$ all have no $J=0$ states.
Apart from the already mentioned triplet $(\ell, N)=(2,3)$, a single $J=0$ state appears in
(3,15), (3,17), (9,5) and (11,5).
As mentioned earlier, energies of these states are linear functions of interaction parameters.
For example, the state of 15 bosons with spin 3 has energy
\begin{equation} 
E=\frac{240}{7}V_2+\frac{2925}{77}V_4+\frac{360}{11}V_6.
\label{eq:15}
\end{equation}

\section{Special Hamiltonians\label{apx:B}}
There are several important special cases of the two-body interaction Hamiltonian (\ref{eq:1}) that allow for 
analytic solutions. 
\subsection{Monopole interaction\label{apx:B1}}
If all matrix elements of the two-body interaction are the same 
\begin{equation}
V^{(m)}_L=1,
\label{eq:16}
\end{equation}
we have the monopole Hamiltonian $H^{(m)}.$
In this case, the pairwise interactions are not sensitive to the type of pairs.
Energies of all many-body states are equal and given by the number of pairs that can be formed from $N$ particles,
\begin{equation}
E^{(m)}=\frac{1}{2}N(N-1).
\label{eq:17}
\end{equation}
\subsection{Pairing}
Pairs of angular momentum $L=0$, whose operators we denote without subscripts $P\equiv P_{00},$ are special from the symmetry perspective and can 
serve as building blocks for a collective $J=0$ pair condensate state.
For a single type of bosons, the commutator of pair operators  is 
\begin{equation}
\left [P,P^\dagger\right ] = \frac{2\ell+1 + 2N}{2\ell+1}.
\end{equation}
 Because of this algebraic property,
the pairing Hamiltonian $H^{(p)}=P^\dagger_{00}P_{00},$ defined with a set of matrix elements  
\begin{equation}
V^{(p)}_L=\delta_{L0}, 
\label{eq:18}
\end{equation}
has eigenvalues
\begin{equation}
E^{(p)}=\frac{(N-\nu)(N+\nu+2\ell-1)}{2(2\ell+1)},
\label{eq:19}
\end{equation}
where $\nu$ is the seniority given by the number of unpaired particles.
The $N-\nu$ particles create a single paired state of zero angular momentum while the $\nu$ unpaired particles
do not participate in the interactions at all, but  add degeneracy and angular momentum to the state.

\subsection{Rotational Hamiltonian}

The square of the angular momentum operator ${\bf J}^2$ can also be constructed using the general form
in eq.~(\ref{eq:1}) plus a one-body term proportional to the number of particles
\begin{equation}
{\bf J}^2=\ell(\ell+1) N+ H^{(j)}.
\label{eq:20}
\end{equation}
Here, the two-body part $H^{(j)}$ is defined by matrix elements
\begin{equation}
V^{(j)}_L=L(L+1)-2\ell(\ell+1).
\label{eq:21}
\end{equation}
The eigenvalues of $H^{(j)}$  are 
\begin{equation}
E^{(j)}=J(J+1)-\ell(\ell+1)N.
\label{eq:22}
\end{equation}

\section{Primed ensemble \label{apx:prime}}

Any Hamiltonian (\ref{eq:1}) always commutes with $H^{(j)}$ and $H^{(m)}$ because angular momentum and particle number
are conserved quantities.  For this reason, the removal of these two components does not change any structure of the wave
functions. Namely, for any $\alpha$ and $\gamma$, the Hamiltonian given by
\begin{equation}
{H'}=H-\alpha H^{(m)}-\gamma  H^{(j)}
\label{eq:xx}
\end{equation}
has wave functions identical to those of $H$; the relationship between energies is 
\begin{equation}
E'_J=E_J-\frac{\alpha}{2} N(N-1)-\gamma \left [ J(J+1)-\ell(\ell+1)N \right ].
\label{eq:yy}
\end{equation}
The transformation in eq.~(\ref{eq:xx}) in terms of matrix elements of (\ref{eq:1}) is 
\begin{equation}
V'_L=V_L-\alpha-\gamma \left(L(L+1)-2\ell(\ell+1) \right). 
\label{eq:39}
\end{equation}
The $\gamma=0$ case amounts to an identical ensemble  where in each realization all energies are shifted by 
a constant.  

We determine $\alpha$ and $\gamma$ from a set of $V_L$ by minimizing the sum $\sum_L(2L+1) V'^2_L,$
which amounts to a fitting of all two-particle states, including their magnetic substates. 
We define the monopole term 
\begin{equation}
M=\frac{\sum (2L+1) V_L}{\sum(2L+1)},
\label{eq:72a}
\end{equation}
and angular momentum term
\begin{equation}
F=\frac{\sum (2L+1) \left(L(L+1)-2\ell(\ell+1) \right) V_L}{\sum(2L+1) \left(L(L+1)-2\ell(\ell+1) \right)}.
\label{eq:73p}
\end{equation}
The minimization procedure determines $\alpha$ and $\gamma$ as
\begin{equation}
\gamma=\frac{3(F-M)}{(2\ell+3)(\ell+2)(2\ell-1)},
\label{eq:37}
\end{equation}
and 
\begin{equation}
\alpha=M-\gamma \ell.
\label{eq:38}
\end{equation}

\bibliography{clboson_f4.bib}

\end{document}